\documentclass[sigconf]{acmart}

\usepackage[shortlabels]{enumitem}

\AtBeginDocument{%
  \providecommand\BibTeX{{%
    \normalfont B\kern-0.5em{\scshape i\kern-0.25em b}\kern-0.8em\TeX}}}

\setcopyright{acmcopyright}
\copyrightyear{2020}
\acmYear{2020}
\acmDOI{xx.xx.xx}

\acmConference[Accepted in ROOTS '20]{ROOTS '20: Reversing and 
  Offensive-oriented Trends Symposium}{November 19--20, 2020}{Vienna, Austria}
\acmBooktitle{ROOTS '20:}
\acmPrice{xx.xx}
\acmISBN{xx.xx.xxx}

\begin{document}

\title{A survey on practical adversarial examples for malware classifiers}

\author{Daniel Park}
\affiliation{%
  \institution{Department of Computer Science\\Rensselaer Polytechnic Institute}
}
\email{parkd5@rpi.edu}

\author{B\"ulent Yener}
\affiliation{%
  \institution{Department of Computer Science\\Rensselaer Polytechnic Institute}
}
\email{yener@cs.rpi.edu}

\renewcommand{\shortauthors}{Daniel Park and Bülent Yener}

\begin{abstract}
  Machine learning based solutions have been very helpful in solving problems that deal with immense amounts of data, such as malware detection and classification. However, deep neural networks have been found to be vulnerable to adversarial examples, or inputs that have been purposefully perturbed to result in an incorrect label. Researchers have shown that this vulnerability can be exploited to create evasive malware samples. However, many proposed attacks do not generate an executable and instead generate a feature vector. To fully understand the impact of adversarial examples on malware detection, we review practical attacks against malware classifiers that generate executable adversarial malware examples. We also discuss current challenges in this area of research, as well as suggestions for improvement and future research directions.
\end{abstract}



\maketitle

\section{Introduction}\label{introduction}
The field of malware detection and classification has grown considerably since the introduction of hash based file signature methods \cite{Morley01}. With malware authors incorporating evasion techniques, such as obfuscation, into their malicious code, detection methods using static and dynamic analysis were and continue to be developed. Increased computational power has also led to many machine learning based solutions being developed alongside these analysis methods and being deployed in commercial products \cite{microsoft-defender2019, avast2020}. 

However, in 2014 Szegedy et. al. showed that deep neural networks (DNNs) are susceptible to adversarial attacks. Grosse et. al. went on to show that this vulnerability also held true for machine learning based malware detectors and classifiers \cite{Grosse2017}. Since this work, there have been many attacks developed against popular machine learning based models such as MalConv \cite{Raff17}, however many of these attacks are not practical. Specifically, many attacks do not generate actual malware and instead generate a feature vector that represents what a possible perturbed malicious file should look like to evade detection. It is unpractical to generate an executable program given a feature vector due to the difficulty of the inverse feature-mapping \cite{Pierazzi2020}. This is to say that the feature extraction process is not uniquely invertible nor is there a guarantee that a found solution would contain the same program logic as the original malware sample.

In this work, we review practical attacks against machine learning based malware classifiers and detectors, or attacks against these ML models that result in executable malware. In Section \ref{background}, we introduce and define adversarial examples and the threat models in which they are considered. Then we review practical adversarial example research in the malware domain in Section \ref{meat}. We offer suggestions for future directions in this field as well as discuss any challenges in Section \ref{discussion}. Lastly, we conclude in Section \ref{conclusion}.

\section{Background}\label{background}
In this section, we begin by briefly discussing popular machine learning methods used in malware classification and detection. Then, we introduce and define adversarial examples and categorize threat models in which the adversarial examples are considered.

\subsection{Machine learning for malware}
With the increasing prevalence of the internet, we have seen an exponential increase in malware and attackers. To exacerbate the problem, malware authors obfuscate their malicious code to impede detection and evade various static and dynamic analysis methods \cite{OKane11}. 
    
The classic approach to detecting malware was to extract a file signature for malicious samples that were found on infected systems and add them to a signature database, also known as signature-based detection \cite{Morley01}. For this approach, the whole sample, as well as subsets of the sample in question, must be searched for known signatures because malicious behavior can be embedded and interleaved in otherwise benign software. However, because signature-based detection relies on capturing a malware sample and then analyzing it to generate a new signature, it can only defend against already known attacks and can only attempt to defend against new or obfuscated malware \cite{Santos09}. Machine learning based approaches have been proposed as a solution to this problem because of their ability to predict labels for new inputs. Machine learning models, such as Support Vector Machines (SVM) and $K$-means clustering, are used in malware classification, detection, and analysis approaches. In the classification problem, we attempt to separate malware samples into predefined malware families. Based on the training data, which consists of labeled malware samples, the learning-based model infers the classification of new malware samples. The detection problem can be seen as a sub-problem of classification. For detection, the learning-based model is used to find or \textit{detect} malware samples when given malicious and benign executables. As detection is a case of binary classification, learning-based detection models can also be called classifiers. Classification and detection are supervised algorithms as the training data is labeled. Machine learning can also be used to augment malware analysis. Non-supervised clustering algorithms can be used to learn new similarities between malware samples \cite{Jang11}. Additionally, we can reason over learning-based models to better understand what makes malware malicious \cite{Arp2014, demetrio2019}. More recently, with increased research in deep learning approaches, researchers have begun to utilize convolutional neural networks to classify and detect malware \cite{Raff17, Nataraj11}.

In Tables \ref{tab:static} and \ref{tab:dynamic}, we provide a brief overview of static and dynamic features that are popularly used in machine learning based malware classifiers. The rest of the section describes proposed models that utilize one or more of the listed features.

\begin{table}
    \caption{In this table, we list static features that are popularly used in machine learning based malware classifiers. We loosely categorize based on the data used to generate each feature.}
    \label{tab:static}
    \begin{tabular}{|c|l|} \hline
    Data     &    Static Feature                                      \\ \hline
    Bytes        &   $\bullet$  Extract $n$-grams from byte sequences  \\
                 &   $\bullet$  Convert bytes to black and white pixels \\
                 &   $\bullet$  Use byte sequence or hex-dump as input \\ \hline
    Opcodes      &   $\bullet$  Opcode frequency vector               \\
                 &   $\bullet$  Extract $n$-grams from disassembly              \\
                 &   $\bullet$  Build Markov chain from opcode sequences             \\
                 &   $\bullet$  Generate control flow graph   \\ \hline
    API Calls    &   $\bullet$  Indicator vector based on data mining                \\
                 &   $\bullet$  Frequency of API call based on data mining          \\ \hline
    System Calls &   $\bullet$  Indicator vector based on disassembly  \\
                 &   $\bullet$  Frequency of system call      \\ \hline
    Environment  &   $\bullet$  Hard-coded network addresses    \\ \hline

    \end{tabular}
\end{table}

\begin{table}
    \caption{In this table, we list dynamic features that are popularly used in machine learning based malware classifiers. Similar to Table \ref{tab:static}, we loosely categorize by the data used to generate each feature.}
    \label{tab:dynamic}
    \begin{tabular}{|c|l|} \hline
    Data     &   Dynamic Feature                                \\ \hline
    Opcodes      &   $\bullet$  $N$-grams extracted from program trace \\ \hline
    API Calls    &   $\bullet$  $N$-grams extracted from program trace \\
                 &   $\bullet$  Frequency of API call during execution \\ \hline
    System Calls &   $\bullet$  $N$-grams based on program trace \\
                 &   $\bullet$  Frequency of system call during execution \\
                 &   $\bullet$  Taint analysis of system information \\ 
                 &   $\bullet$  Behavioral profiles using system interaction \\\hline
    Environment  &   $\bullet$  Data transfers \\ 
                 &   $\bullet$  Network traffic and communication \\
                 &   $\bullet$  File and registry edits, creation, and deletion \\ 
                 &   $\bullet$  New processes spawned during execution \\ \hline

    \end{tabular}
\end{table}

\subsubsection{Static Features}
$N$-grams are currently a popular feature used for the classification and detection of malware. Kolter and Maloof proposed extracting the most relevant $n$-grams of byte codes from PE malware for classification using various machine learning models, including naive bayes classifiers, decision trees, support vector machines (SVM), and boosted models \cite{Kolter06}. They found that in addition to classification, their model could be used for malware detection.

McBoost was introduced as a tool to quickly analyze large amounts of binaries when searching for malware and utilized a three step process \cite{Perdisci08}. The first step is to detect packers using an ensemble of a heuristic-based classifiers and two different n-grams based classifiers. If a packer is detected, the binary is unpacked using QEMU and dynamic analysis. Lastly, a separate n-gram classifier is used to detect malware that should be forwarded for additional analysis.

Santos et. al. proposed the use of $n$-grams as an alternative to file signatures based methods in 2009. In doing so, they showed that machine learning, specifically a k-nearest neighbors model, and $n$-grams can successfully be used to detect new malware samples \cite{Santos09}. $N$-grams have also been used together with dynamic features to incorporate multiple views of malware simultaneously without actually conducting dynamic analysis \cite{Blake12}. Recent work such as solutions proposed for the Kaggle Microsoft Malware Challenge \cite{Ronen18, saynotooverfitting}, demonstrate continued usage of both byte and opcode $n$-grams with classification accuracies of almost $100\%$. 

Similar to using sequences of opcodes to generate $n$-grams, Markov chains can be extracted from the opcode trace from a program. Such a Markov chain uses unique opcodes as its states and shows the transition probabilities from one opcode to another. Anderson et. al. used the similarity between programs' Markov chains as one feature for malware detection. Similarly, Runwal et. al. \cite{Runwal2012} proposed using graph similarity between Markov chains to detect embedded malware and Shafiq et. al. \cite{Shafiq2008} proposed measuring the entropy using Markov chains to detect malware. 

Jang et. al. presented BitShred, a tool for malware triage and analysis \cite{Jang11}. BitShred hashes the $n$-gram fingerprints extracted from samples using locality sensitive hashing to reduce the dimensionality of the feature space. The hashes are used in a k-nearest neighbors model to cluster the malware samples. Additionally, the authors showed that BitShred can be used to improve previous malware detection and classification models. For example, the authors showed that BitShred can be used to hash dynamic features, such as the behavior profiles generated in Bayer et. al. \cite{Bayer09}, to reduce the dimensionality of the feature space.

Drebin conducts large-scale static analysis of Android software to extract features such as hardware usage and requests, permissions, API calls, and networking information \cite{Arp2014}. These features are used to map the sample to a joint vector space by generating an binary indicator vector. These binary indicator vectors are used as input to a SVM, which labels a sample as benign or malicious. Importantly, Drebin takes advantage of the simplicity of their model to attribute the model's decisions to specific features. This makes Drebin more explainable than malware classifiers and detectors based on complex architectures, such as convolutional neural networks.
    
Nataraj et. al. proposed using malware images (black and white image representations of binaries) to detect malware \cite{Nataraj11}. Since then, researchers and commercial antivirus have used malware images to detect malware with high accuracy \cite{saynotooverfitting, Ghouti20, Kamundala18}. Nataraj et. al. used the malware images to create a feature vector that would be used as input to a support vector machine. However, recent work has also shown that it is effective to use the raw images as input to a convolutional neural network \cite{Yan18,Kornish18,Kalash18,Kamundala18}.

Research has also been done in developing static features that give some insight into how the program behaves during run-time using control flow graphs. This research mainly revolves around constructing a control flow graph and using graph matching techniques to detect malware \cite{Cesare2011, Bruschi2006}. Ma et. al. take a similar approach in using control flow graphs, but extracts a sequences of API calls in an attempt to mimic dynamic analysis \cite{Ma2019}.

Raff et. al. takes a different approach and propose a convolutional neural network (CNN) model that takes in the whole binary as input \cite{Raff17}. In particular, the proposed model MalConv looks at the raw bytes of the file to determine maliciousness. MalConv borrows from neural network research in learning higher level representations from long sequences and relies on CNN's ability to capture high level local invariances. MalConv works by extracting the $k$ bytes of a file. These $k$ bytes are padded with $0xff$ bytes to create a feature vector of size $d$. If $k > d$, the first $d$ bytes of the file are used without any additional padding. This $d$-length vector is mapped to a fixed length feature vector by an embedding that is jointly learned with the CNN.  
        
\subsubsection{Dynamic Features}
Dynamic analysis is a technique for analyzing a binary by running it in a live environment. This environment is usually a secure sandbox or test environment, such as CWSandbox \cite{cwsandbox} and Cuckoo Sandbox \cite{cuckoo}, as to keep the host machine safe. Generally, these environments are heavily instrumented in order to keep record of executed and loaded code as well as any changes made to internal files, directories, and settings. These recorded features are called dynamic features.
        
The most general method of extracting dynamic features is to record the frequency and sequencing of system and API calls \cite{Afonso2014, Dim2016}. For example, Accessminer records the systems calls trace during dynamic analysis and generates an $n$-grams representation of each sample \cite{Lanzi10}. Accessminer labels a sample as malware if it contains more multiple instances of "malicious" $n$-grams with respect to some predefined threshold. Another benefit of dynamic analysis is that network traffic and communication can be captured and analyzed, as proposed in Taintdroid \cite{Enck2010}. These features can also be used to generate different representations of malware for additional features or to reduce dimensionality.

Bailey et. al. presented a dynamic analysis tool for automated classification and analysis of malware that used dynamic analysis to record new processes that were spawned, any files that were modified, any registry keys that were modified, as well as network access and usage \cite{Bailey07}. These recorded features were used to create a malware fingerprint that focuses on state changes instead of code sequences. These dynamic features are used to create a hierarchical clustering of malware samples using a normalized compression distance metric. 

Rieck et. al. uses CWSandbox to conducts dynamic analysis similar to the work by Bailey et. al., however, extracts features using strings from the resulting text report \cite{Rieck08}. The string frequencies are used with a SVM to classify malware samples. The authors also show that their method can be extended to malware detection by introducing a new "unknown" class without introducing benign samples into the training set.

Bayer et. al. extended upon previous work by using taint analysis to learn how information from the operating system is used by the executable \cite{Bayer09}. Additionally, the proposed method uses an abstraction of operating systems objects and operations to create a behavioral profile. The authors argue that the abstractions are more robust against evasion due to being able to abstract away or reason about the program without bogus system calls. The extracted behavioral profiles are then used with a clustering algorithm based on locality sensitive hashing to classify malware samples.

The behavior of a program can also be modeled into graphs as in the work of Kolbitsch et. al. \cite{Kolbitsch09}. The authors extended upon Malspecs \cite{Christodorescu07} and generated behavior graphs of programs using system calls. Each behavior graph is a directed acyclic graph where the nodes are system calls and the directed edges denote information flow. Malware samples are detected using graph matching and similarity metrics against already known malware samples.

\subsection{Adversarial examples}
The notion of adversarial examples was first introduced in \cite{Szegedy2014} and expanded upon in \cite{Goodfellow14}. Assume $f$ is the target classifier that an adversary plans to attack. This classifier can be represented as a function $f(\cdot)$ that takes an input and assigns it a label. An adversarial example $x'$ targeting $f$ is generated by perturbing an original input $x$ with $\delta$ so that $f(x) \ne f(x')$.
$$ x' = x + \delta ,$$
$$ f(x') \ne f(x) $$

Additionally, the perturbation is generally bounded by some value $\epsilon$ using an $l_{2}$, $l_{1}$, or $l_{\infty}$ norm.
$$ || x' - x ||_{p}  < \epsilon $$
In the natural image or sound domain, this bound is used to ensure that the perturbations are imperceptible to a discriminator, such as the human eye or ear.

There are many ways to find $\delta$, the most popular being the Fast Gradient Sign Method  \cite{Goodfellow14} and the Carlini-Wagner (C\&W) Attack \cite{Carlini16} for white box models and substitute model attacks \cite{Papernot2016} for black box models. Most attacks will use the gradient of the loss function with respect to the input to find the direction in which the input must be perturbed for the wanted change to the output. This direction is then used to find $\delta$. We briefly discuss these and other attacks in the Appendix.

\subsubsection{Threat models}
A threat model is a clear definition of the adversary that is considered in a study, including all capabilities and knowledge of the adversary. In this section, we define the white box and black box threat models that are popularly used in the machine learning domain. The threat models are made up of three parts: \textit{threat vector and surface}, \textit{knowledge}, and \textit{capabilities}.

\textbf{Threat Vector and Surface:} The threat vector and surface correspond to the means in which the adversary interacts with the model under analysis. The threat vector is the allowable input space and locations that the adversary can use to attack the model. The threat surface, or attack surface, is the collection of all such threat vectors. In this case, the threat vector and surface consist of the input and output of the machine learning model. However, the adversary's access to these surfaces is further constrained by their knowledge and capabilities.
    
\textbf{Knowledge:} The adversary's knowledge represents what we assume the adversary knows about the target model. This knowledge is then used by the adversary to construct and mount an attack. In adversarial machine learning, the adversary's knowledge can be generalized into white and black box models. 
    
In the white box model, the adversary is assumed to have complete knowledge of the system. Thus, we assume the adversary has  complete access to the target machine learning model (with weights and parameters) as well as the data used to train the model. In the black box model, the adversary is assumed to only have access to the input and output of the model. Thus, the adversary has no knowledge of the internals or training process of the model (e.g., features extracted from executables and gradient information). An adversary can also be modeled as gray box. In a gray box model, the adversary has access to the input and output of the model and more. However, as gray box covers the entire spectrum between white and black box, it must be carefully defined. 

The works reviewed in this study do not agree exactly on the adversary's knowledge. Specifically, some works may assume the adversary additionally has access to malware source code, while others do not. In Section \ref{meat}, each works' deviations from this general definition of adversarial knowledge will be clearly denoted.
    
\textbf{Capabilities:} An adversary's capabilities corresponds to the abilities of the adversary and the types of attacks that can be mounted. In the case of adversarial examples, we may specify an attack algorithm that the adversary will use. Furthermore, in the case of adversarial examples in the malware domain, the adversary's capabilities are limited by their knowledge. For example, with access to malware source code, an adversary can easily apply specific transformations at compile time. However, this becomes more difficult without the source code.

\subsubsection{Adversarial malware examples} \label{ame}
Most adversarial example research is conducted using natural image datasets, such as MNIST, CIFAR10, and ImageNet. However, it is necessary to consider the set of allowable perturbations that preserve functionality of adversarial malware examples.

For natural images, the pixel values are perturbed to generate an adversarial example. Any negative or positive change to a pixel value will result in a slightly altered image as long as the resulting pixel is between 0 and 255. Executable programs can be represented in a similar way. Each byte of the binary, by definition, is between \textit{0x00} and \textit{0xff}. Each byte's hex representation can be translated to its decimal equivalent (between 0 and 255). In this state, a byte and pixel can be perturbed using the same methods. However, an arbitrary perturbation to a byte may not result in a valid executable because executable programs exist in a discrete space. Consider the simple case of altering one byte of an executable. If the byte comes from the \textit{.text} section of an ELF, the new altered byte may break functionality of the program by changing function arguments or resulting in bad instructions. For this reason, applying adversarial example techniques to the malware domain requires special care in the binary's construction. Most importantly, an adversarial malware example must contain the same malicious program logic and functionality as the original. 

Adversarial malware examples are an immediate threat as they are evasive and malicious executables that can take advantage of many commercial antivirus software's persistent vulnerability to obfuscation and mutation \cite{Quarta2018}. This differentiates practical adversarial malware examples from an adversarial feature vector. While an adversarial feature vector also evades detection or classification, there is no immediate threat. Pierazzi et. al argue that generating an executable given an adversarial feature vector is difficult and call this the \textit{inverse feature-mapping} problem. There is no unique solution to the inverse feature-mapping problem. In the simple case of an $n$-gram classifier, the addition of an $n$-gram can be done in multiple ways. However, they are not all guaranteed to result in an executable that contains the same program logic or executability as the original malware sample. This problem becomes more difficult when dealing with black box models, where the attacker has no knowledge of the classifier's input and internals. Pierazzi et. al. explain that there are two ways practical adversarial malware examples circumvent this: (1) A gradient-driven approach where the code perturbations' effect on the gradient is approximated and used to follow the direction of the gradient and (2) a problem-driven approach where mutations are first applied randomly before taking on an evolutionary approach.

\begin{table*}
    \caption{In this table, we summarize practical adversarial malware example algorithms by showing (1) if the work was evaluated with static features, (2) if the work was evaluated with dynamic features, (3) the target learning-based models in the evaluation, (4) available transformations in the attack, and (5) whether the attack is gradient-driven or problem-driven. We note a general trend in the use of obfuscation and further discuss this in Section \ref{discussion}. }
    \label{tab:tax}
    \begin{tabular}{|c|c|c|p{.3\textwidth}|p{.13\textwidth}|p{.125\textwidth}|} \hline
    Attack & Static & Dynamic & Target model & Transformation & Approach \\ \hline
     GADGET \cite{rosenberg2017}              & \checkmark & \checkmark & Ensemble of custom machine learning and deep learning models & add API calls & gradient-driven \\ \hline
     Anderson et. al. \cite{anderson2018}     & \checkmark & & Gradient boosted decision trees & Edit byte features and PE metadata  & problem-driven \\ \hline
     Kolosnjaji et. al. \cite{kolosnjaji2018} & \checkmark & & MalConv \cite{Raff17} & Edit padding bytes & gradient-driven \\ \hline
     Kruek et. al. \cite{kreuk2019}           & \checkmark & & MalConv \cite{Raff17} & Edit padding bytes & gradient-driven \\ \hline
     Demetrio et. al. \cite{demetrio2019}     & \checkmark & & MalConv \cite{Raff17} & Edit PE header bytes & gradient-driven \\ \hline
     Park et. al. \cite{park2019}             & \checkmark & & Custom CNNs and gradient boosted decision trees & semantic $NOP$ insertion & gradient-driven \\ \hline
     Song et. al. \cite{song2020}             & \checkmark & \checkmark & Commercial antivirus & Edit byte features and PE metadata, and instruction substitution & problem-driven \\ \hline
     Yang et. al. \cite{Yang2017}             & \checkmark &  &  AppContext \cite{Yang2015} and Drebin \cite{Arp2014} & Mutates and transplants context features \cite{Yang2015} & problem-driven \\ \hline
     Kucuk et. al. \cite{kucuk2020}           & \checkmark & \checkmark  & Two custom classifiers using static features and one custom classifier using dynamic features & Control flow obfuscation, bogus code blocks, and add API calls & problem-driven   \\ \hline
     Pierazzi et. al. \cite{Pierazzi2020}     & \checkmark &  & Drebin \cite{Arp2014} & Bogus code blocks and opaque constructs & problem-driven \\ \hline
     HideNoSeek \cite{Fass2019}               & \checkmark &  & Custom Bayesian classifier & JavaScript AST transforms & problem-driven \\ \hline
             
    \end{tabular}
\end{table*}

\section{Practical attacks}\label{meat}
In this section, we review practical attacks in the adversarial malware example literature, or attacks that result in executable binaries. In Table \ref{tab:tax}, we give an overview of the practical attacks in this work recording (1) if the work was evaluated against malware classifiers that use static features (2) if the work was evaluated against malware classifiers that use dynamic features, (3) the target models in their evaluation, (4) available transformations in the attack, and (5) whether the approach is gradient-driven or problem-driven. 

We use terminology from \cite{Pierazzi2020} and organize our review into gradient-driven and problem-drive approaches as defined in Section \ref{ame}. For both approaches, we further organize the literature into attacks that mainly edit bytes and metadata in Sections \ref{bytes1} \& \ref{bytes2} and attacks that utilize code transformations in Sections \ref{transform1} \& \ref{transform2}.

\subsection{Gradient-driven approaches} \label{gradient}
In this section, we review gradient-driven approaches for generating adversarial malware examples. We further organize the review using the attacks' available transformations.

\subsubsection{Editing bytes and metadata} \label{bytes1}
A popular method for creating practical adversarial malware examples is to add or alter bytes in unused space in the binary. Additionally, this can be done in the header to change header metadata without affecting functionality. In this section, we will review proposed attacks that use this type of transformation. Because these attacks focus on unused or "unimportant" (for execution) bytes, they do not require source code for generating their evasive malware samples. However, with the exception of GADGET \cite{rosenberg2017}, these attacks are still white box attacks as they require complete access to the target model to compute gradients.

In 2018, Rosenberg et. al. proposed GADGET, a software framework to transform PE malware into evasive variants taking advantage of the transferability property of adversarial examples between DNNs \cite{rosenberg2017}. The proposed attack assumes a black box threat model with no access to the malware source code. However, the attack assumes the target model takes a sequence of API calls as input. To generate adversarial examples, GADGET constructs a surrogate or substitute model that is trained with Jacobian-based dataset augmentation, introduced by Papernot et. al. as an attack against natural image classifiers \cite{Papernot2016}. The dataset augmentation creates synthetic inputs that help the substitute model better approximate the target black box model's decision boundaries. This increases the probability of the transferability of the attack as both the substitute and target model will have learned similar distributions. Once the substitute model is trained, adversarial malware examples are generated by adding dummy API calls to the original malware's API call sequence. The authors call these dummy API calls semantic \textit{nops} as the chosen API call or their corresponding arguments have no affect on the original program logic. It is important to note that the authors only add API calls, as removing an API call can break the functionality of the program. Let us say that the original API call sequence is an array $w_0$ where each index $j \in [0, n]$ contains an API call. Each iteration $i$ of this process returns a new array $w_i$. At iteration $i$, an API call $d$ is added to $w_{i-1}$ at some index $j$ that pushes it towards the direction indicated by the gradient as the most impactful for the substitute model's decision. This results in $w_i$ where $w_i[j] = d$ and $w_i[j+1:] = w_{i-1}[j:]$ because all API calls in the previous sequence after index $j$ are essentially "pushed back". This method of perturbing the input by adding dummy API calls ensures that functionality is not broken. To generate the actual executable from this adversarial API call sequence, GADGET implements a wrapper that hooks all API calls. The hooks call the original APIs as well as dummy APIs as necessary from the adversarial API call sequence. These hooks ensure that the resulting adversarial malware example maintains the functionality and behavior of the original sample, as the original sample is being executed in a sense. GADGET was evaluated against Custom models including variants of logistic regression, recurrent neural networks (RNN), fully connected deep neural networks (DNN), convolutional neural networks (CNN), SVM, boosted decision trees, and random forest classifiers. The authors also showed that their attack produces malware that is able to evade classifiers that use static features, such as printable strings.

Kolosnjaji et. al. proposed a white box attack against MalConv that generated adversarial PE malware examples by iteratively manipulating padding bytes at the end of the file \cite{kolosnjaji2018}. Although the authors note that bytes at any location in the PE can be altered, it requires precise knowledge of the file architecture as a simple change can break file integrity. For this reason, the proposed attack focused only on byte appending. A challenge faced by the authors was the non-differentiability of MalConv due to its embedding layer. To circumvent this, the authors proposed computing the gradient of the objective function with respect to the embedding representation $z$ instead of the input. Each padding byte is replaced with an embedded byte $m$ that is closest to the line $g(\eta) = z + \eta n$ where $n$ is the normalized gradient direction. However, if $m$'s projection on the line $g(\eta)$ is not aligned with $n$, the next closest embedded byte is selected. By only altering the padding at the end of the file, the proposed attack does not change the program logic nor the functionality of the original malware sample. However, this also limits the total number of perturbations allowed by the attack. As explained in Section \ref{background}, MalConv extracts up to $d$ bytes from a binary. If the size of the binary is less than $d$, the extracted $k$ bytes have $(d-k)$ \textit{0xff} padding bytes appended to it. This means that the proposed attack is limited by the size of the original malware sample. 

Kruek et. al. \cite{kreuk2019} extended the work of Kolosnjaji et. al. by proposing a method for reconstructing the PE malware sample given the adversarial example's embedding. The authors found that reconstructing bytes from the perturbed embedding $z*$ is often non-trivial as $z*$ can lose resemblance to embeddings $z \in Z$ used to learn $M$, the function mapping padding bytes to embedding bytes. Thus, they presented a novel loss function to ensure that perturbed embeddings $z*$ will be close to an actual embedding in $M$. This is done by introducing a distance term in the loss function between generated embeddings and $M$. 

Demetrio et. al. proposed \textit{feature attribution} as a explainable machine learning algorithm to understand decisions made by machine learning models \cite{demetrio2019}. Feature attribution was based off of a technique called \textit{integrated gradients} introduced in 2017 by Sundararajan et. al \cite{sundararajan2017}. Given the target model $f$, an input $x$, and a baseline $x'$, integrated gradients compute the attribution for the $i$th feature of $x$ as
$$ IG_i(x) = (x_i - x_i') \int_0^1 \frac{\partial f(x' + \alpha (x-x'))}{\partial x_i} d\alpha$$
As this integral is computed on all points in the line between $x$ and $x'$, each point should contribute to $f$'s classification of $x$ as long as $x'$ is a \textit{good} baseline. The authors approximate this integral using summations, pulling from the suggestions in \cite{sundararajan2017}. It is important to note that these contributions are calculated with respect to the chosen baseline $x'$. The authors selected an empty file to be the baseline for the proposed feature attribution technique. Another option for the baseline was a file with only zero bytes. However, this option was labeled malicious with a $20\%$ probability by MalConv, going against baseline constraints laid out in \cite{sundararajan2017}. Using feature attribution, Demetrio et. al. observed the attribution to each byte of input executables and found that MalConv heavily weighs the PE header section of binaries. The authors exploited this and presented a white box attack against MalConv that only alters bytes in the malware sample's header. This attack used the same algorithm presented in \cite{kolosnjaji2018} but perturbed unused and editable bytes inside the header instead of padding at the end of the file.

\subsubsection{Code transformations} \label{transform1}
Many of the works above note that the proposed methods can be used to alter the \textit{.text} section of malicious binaries as long as the program's functionality and malicious behavior is not altered. The following attacks make use of obfuscation techniques to alter the \textit{.text} section.

Park et. al. proposed a white box attack that utilized \textit{semantic nops}, such as \textit{mov eax, eax} in x86 assembly, to create adversarial PE malware examples \cite{park2019}. The authors attacked convolutional neural networks that used the image representation of an executable \cite{Nataraj11} as input. The image representation of an executable treats each byte as a pixel and uses the byte's decimal value as the pixel value. The proposed attack has two steps. First, an adversarial example is generated using FGSM. This adversarial example is an image and may not have the same functionality or malicious behavior as the original malware sample. In the second step, the original malware sample and the generated adversarial image are used as input to a dynamic programming algorithm that inserts \textit{semantic nops} using LLVM passes. Similar to how API calls are added to resemble the generated adversarial feature vector in \cite{rosenberg2017}, the dynamic programming algorithm adds \textit{semantic nops} in a way such that the resulting malware sample's image representation resembles the generated adversarial image from step 1. The authors went on to show that this attack can be used against a black box model because of the transferability property of adversarial examples and perturbations \cite{Papernot2016, Moosavidezfooli16}. Using a simple 2-layer CNN as a substitute model, the authors generated adversarial malware examples that also evaded black box models, one of which being a gradient boosted decision tree using byte-level features. The authors also mention that their attack works best given the malware's source code. However, in the absence of source code, binary translation and rewriting techniques can be used to insert the necessary \textit{semantic nops}. It is important to note that introducing these techniques also introduces artifacts from binary lifting process.

\subsection{Problem-driven approaches}
In this section, we review adversarial malware example algorithms that take problem-driven approaches. Similar to Section \ref{gradient}, we further organize the review using the attacks' available transformations. Problem-driven approaches do not require white box access to the target for gradient information. As such, the following approaches are black box attacks.

\subsubsection{Editing bytes and metadata} \label{bytes2}

Anderson et. al. proposed a particularly interesting attack in which a reinforcement learning (RL) agent is equipped with a set of PE functionality preserving operations \cite{anderson2018}. The RL agent is rewarded for actions that produce malware that evades detection. Through this game, the agent learns a policy for creating evasive malware. The proposed attack makes use of the following actions that do not change the original program logic:
\begin{itemize}
    \item Adding functions to the import table that are never used.
    \item Changing section names.
    \item Creating new but unused sections.
    \item Adding bytes to unused space in sections.
    \item Removing signer information.
    \item Alter debugging information.
    \item Packing or unpacking the binary.
    \item Modifying the header.
\end{itemize}
Using these actions, the RL agent is able to alter features such as PE metadata, human readable strings, and byte histograms. After up to 50,000 mutations during the training phase, the RL agent is evaluated against a gradient boosted decision tree model, shown to be successful in classifying malware \cite{saynotooverfitting}. The authors note that their adversarial examples should be functional by construction. However, they found that their attack breaks functionality in certain Windows PE's that make use of less common uses of the file format or obfuscation tricks that violate the PE standard. The authors claim that this can simply be fixed by ensuring the original malware samples can be correctly parsed by binary instrumentation frameworks.

Song et. al. took a different approach in generating adversarial malware examples \cite{song2020}. The proposed attack randomly generates a sequence of \textit{macro-actions} and applies them to the original PE malware sample. This is repeated until the resulting transformed malware evades detection. Once the malware sample is evasive, unnecessary macro-actions are removed from the sequence of macro-actions applied to it. This is done to minimize the probability of accidentally breaking functionality due to some obfuscation tricks. The remaining macro-actions are then broken down into \textit{micro-actions} for a finer detailed trace of transformations leading to the adversarial malware sample. We suggest the reader peruse the original paper for greater detail on each macro and micro action, however, we briefly describe them here. Macro-actions consist of the following:

\vspace{2mm}
\begin{itemize}[-,nosep]
    \item Append bytes to the end of the binary.
    \item Append bytes to unused space at the end of a section.
    \item Add a new section.
    \item Rename a section.
    \item Zero out signed certificate.
    \item Remove debugging information.
    \item Zero out the checksum value in the header.
    \item Substitute instructions with semantically equivalent instructions.
\end{itemize}
\vspace{2mm}

Some of these macro-actions can be broken down to a sequence of smaller actions, named micro-actions. For example, the action of appending bytes can be broken down to a sequence of adding one byte at a time. The authors claim that by breaking down each macro-action, it is possible to gain insights into why a particular action caused evasion. Instead of utilizing adversarial example generation algorithms such as FGSM or the C\&W attack, the proposed method instead seeks to provide a more explainable attack against machine learning models. This method was evaluated against commercial antivirus and was also found to be effective against classifiers that incorporate both static and dynamic analysis.

\subsubsection{Code transformations} \label{transform2}
Yang et. al. proposed two attacks to construct Android malware samples to evade detection by machine learning models, but did not use machine learning algorithms \cite{Yang2017}. Instead of targeting misclassification, the proposed \textit{evolution attack} focuses on mimicking the natural evolution of Android malware based on mutating contextual features (made up of temporal features, locale features, and dependency features) \cite{Yang2015}. This is done by automating these mutation strategies through an obfuscation tool OCTOPUS and employing them on a large scale to identify "blind spots" on target classifiers. Malware families are organized into phylogenetic evolutionary trees \cite{Chen2006} to analyze commonly shared features and divergent features within the family. Each feature mutation is then ranked by feasibility and frequency, and sorted. The top $x$ mutations are then used to generate new malware variants. The authors also proposed a \textit{feature confusion attack} to complement the evolution attack. The goal of the feature confusion attack is to modify the malware sample such that certain features are similar to those of benign samples. The attack begins by collecting a set of \textit{confusing features}, or a set of features that both malware and benign samples share. For each feature in the \textit{confusing feature} set, the number of benign and malicious samples containing that feature is recorded. If there are more benign samples, that feature is added to the target features list. The attack then mutates malware samples to include the found target features to increase probability of evasion. The proposed method was evaluated against Android learning-based malware classifiers AppContext \cite{Yang2015} and Drebin \cite{Arp2014}. It is important to note that while the attack does not require white box access to the target model, it does assume (1) malware source code and (2) knowledge of features used by the model.

Kucuk et. al. argued that adversarial malware examples must evade both static and dynamic machine learning based classifiers \cite{kucuk2020}. As such, they proposed an attack for PE malware utilizing bogus control flow obfuscation and API obfuscation to evade detection by models using both static and dynamic features. The applied control flow obfuscation is based off of the LLVM-Obfuscator \cite{Junod15}. LLVM-Obfuscator alters the control flow of a program at the LLVM-IR level by utilizing opaque predicates and never-executed bogus basic blocks with arbitrary instructions. Using differential analysis, the authors find the optimal control flow obfuscation and bogus basic blocks to generate an adversarial malware example. This perturbs static features, such as $n$-grams, opcode frequency, and imported API calls. The attack uses a genetic algorithm minimizing the Kullback-Leibler (KL) divergence between the frequency feature vectors of the desired target class and the adversarial malware sample. To evade a dynamic API call based malware classifier, the authors use the same genetic algorithm to determine which API calls must be obfuscated and then obfuscates them using the techniques laid out in \cite{suenga2009}. Additionally, the same genetic algorithm is used again to determine additional API call sequences that should be added to the original malware sample, similar to the approach taken by \cite{rosenberg2017}.

Pierazzi et. al. proposed a black box attack targeting the Android malware classifier Drebin \cite{Pierazzi2020}. The authors proposed a problem-space approach that repeatedly inserts benign code blocks using opaque predicates to change features extracted by Drebin. These benign code blocks are initialized before the attack by analyzing samples in the training set for code sequences that contribute to a negative or benign label. The attack is bounded by a feasibility check to avoid excessive transformations, which may lead to increased suspicion. Additionally, the code blocks are inserted using FlowDroid \cite{Arzt2014} and Soot \cite{soot} to minimize side-effects or artifacts.

HideNoSeek differs from other attacks that apply code transformations in that it attempts to hide malicious JavaScript by transforming the abstract syntax tree (AST) to appear benign \cite{Fass2019}. The attack begins by building ASTs of malicious and benign files to detect sub-ASTs or sub-graphs that are shared between the two classes. To create the adversarial example, HideNoSeek utilizes randomization, data obfuscation, and opaque constructs to insert benign-appearing sub-ASTs. The attack can also rewrite existing ASTs to appear benign. These attacks were conducted in a black box model against custom classifiers based on Zozzle, a Bayesian classifier that uses features extracted from JavaScript ASTs \cite{Curtsinger2011}.

\section{Discussion}\label{discussion}
In this section, we discuss challenges in practical adversarial malware sample research as well as possible research directions.
\subsection{Challenges}
First, it should also be noted that we are in no way diminishing or downplaying the contributions of adversarial example research in the malware domain that do not result in an executable malware sample. However, we believe that extending or including a discussion about possible ways to extend the attack to result in an executable malware sample is necessary to better frame the proposed attack in a realistic adversarial environment. As adversarial example research is growing at a rapid pace, it is necessary to fully understand how these attacks can transition to the malware detection and cybersecurity field. A fully developed or proof of concept attack would also aid in the development of models robust against adversarial malware samples. 

\subsubsection{Threat models.}
One challenge in this area of research is inconsistency in threat models. We believe it is necessary to clearly define the threat model considered in each study to better understand the limitations of the attacks as well as any assumptions made by the authors. In addition to the general white box and black box threat models used in the adversarial example literature, we recommend including (1) assumptions on source code availability and (2) feasibility of attack due to time or computational constraints on the adversary. Similar to the work of Papernot et. al. \cite{Papernot2016}, it would be interesting to see the effects of varying the adversary's resources, e.g., limiting the allowed number of queries to the target model or incurring a cost for each iteration of the adversary's attack. 

\subsubsection{Establishing baselines.}
Another challenge is in establishing baselines and ground truth. There is no consistent dataset nor are there consistent (ML or commercial) malware classifiers throughout the reviewed papers. Although all works considered in this survey boast high evasion rates against top classifiers, we are unable to fairly evaluate them against each other. Having consistency between proposed attacks and their experimental evaluation would allow for better comparisons between the attacks. However, maintaining consistent datasets and malware classifiers and conducting a fair evaluation both pose their own challenges as shown in \cite{VanDerKouwe2020}. This would also help extend the evaluation set forth by Quarta et. al., who used their framework crAVe to show that simply obfuscating or mutating malware samples can be enough to evade detection as not all anti-virus software conduct some form of dynamic analysis.

\textit{Dataset: }Although transforming old malware to be evasive does show a vulnerability in malware detection, exclusion of more recent malware samples poses a risk in concept drift. For example, if malware changes drastically to target new platforms, old malware datasets may not correctly reflect malicious features and behavior. The same can be said for benign program samples. Traditionally, benign samples are scraped from fresh installations of an operating system. However, it is unclear whether these pre-installed programs are reflective of programs that a user downloads and/or scans for malicious behavior.

\textit{Malware classifiers: }It is currently unclear which malware classifier is the best for evaluating an attack. As Song et. al. stated, it is also unrealistic to assume any prior knowledge of the model. We do not believe there currently is nor will there be a consistent malware detection model baseline as research in this area is still growing. However, we do suggest that future work evaluate their attacks against multiple classifiers under a black box threat model. This would be helpful in understanding the attack's transferability between various detection models that use different features in their decision-making process.

\subsection{Possible research directions}
In this section, we will briefly discuss currently open research questions. 

\subsubsection{Defending against practical adversarial malware examples. } Some research has already been done in evaluating the use of adversarial training in the malware domain \cite{aldujaili2018, Li2020}. However, robust machine learning research includes many other defense strategies such as smoothing \cite{cohen2019} and  randomization \cite{pinot2020}. It is unclear whether these approaches would transfer and defend against adversarial malware examples.

\subsubsection{Relationships between obfuscation and adversarial examples.} Obfuscation and adversarial examples share a common goal: evade detection. Additionally, a majority of the practical adversarial malware example algorithms incorporated popular obfuscation strategies into their attack. One possible research problem is evaluating the feasibility of using more advanced obfuscation methods, such as virtualization, for adversarial example generation. It is also currently unclear what the benefit of adversarial malware examples is when compared against more traditional malware evasion techniques, such as  those summarized in Bulazel et. al. \cite{bulazel2017}. It would also be interesting to extend upon the work of Song et. al. \cite{song2020} and Demetrio et. al. \cite{demetrio2019} in the explainability of adversarial malware examples and use this to further develop evasive transformations.

\subsubsection{Integration of static and dynamic analysis techniques.} Many of the reviewed works assume that no advanced analysis is done on the malware samples prior to being tested. However, this does not always have to be the case. For example, a pre-processing step can be used to deobfuscate the evasive malware samples produced by \cite{park2019} and \cite{kucuk2020} using a deobfuscation framework such as SATURN \cite{garba2019}. It would be interesting to see future work in attack and defense that considers using classification and detection pipelines instead of a sole machine learning model or commercial antivirus product.

\subsection{Other Survey and systematization of knowledge papers} In this section, we provide other survey and systematization of knowledge papers that cover related topics.

Yuan et. al. survey adversarial attacks and defenses for deep learning \cite{Yuan2019}. They also provide applications and problem domains in which adversarial attacks can be used. Similar to this work, Maiorca et. al. provide a survey on adversarial attacks against machine learning based PDF malware detection systems \cite{Maiorca2019}. Bulazel and Yener survey dynamic malware analysis evasion and mitigation strategies \cite{bulazel2017}. Ye et. al. survey the application of data mining techniques to malware detection \cite{Ye2017}. Ucci et. al. provides a survey on malware analysis using machine learning \cite{Ucci2019}. Lastly, van der Kuowe et. al. survey popular benchmarking flaws that must be considered to fairly and accurately evaluate security research.

\section{Conclusion}\label{conclusion}
We have presented a survey of practical adversarial examples in the malware domain. The study of adversarial examples and their affect on the cybersecurity field is incredibly important as machine learning based solutions begin to be adopted in both industry and academia. We hope that this survey will provide to be useful in future research in this field.

\begin{acks}
We would like to thank Adrian Dabrowski and the anonymous reviewers for their helpful comments and suggestions for improving the paper.
\end{acks}

\bibliographystyle{ACM-Reference-Format}
\bibliography{acmart}
\appendix
\section{Adversarial examples}\label{appendix}
In this section, we gently present high level descriptions of popular white and black box attacks against DNN's. These descriptions are included to showcase the expansiveness of this research area and to hopefully spark novel applications to the malware domain.

\subsection{Preliminaries}
Many of the papers in this field use slightly differing terminology and variables to denote the same thing. In this section, we will provide definitions that will be used for the rest of the document for consistency.
\begin{table}[h]
\centering
\begin{tabular}{|c|p{6cm}|} 
     \hline
     $F(x, \theta) = y$ &  Neural network with parameters $\theta$ that accepts input $x\in\mathbb{R}^n$ and produces output $y\in\mathbb{R}^m$. Often, $\theta$ will be omitted if it is fixed. \\ \hline
     $Z_F(x)$ & Similar to $F(x)$ but does not include the softmax layer (i.e., the logits). $F(x) = \text{softmax}(Z_F(x))$ . \\ \hline
     $C_F(x) = l$ & Assigned label for $x$ by neural network $F$, i.e. $arg max_i \ F(x)_i$. $F$ will be omitted if the network in question is clear.\\ \hline
     $C^*(x)$ & The correct label for $x$. \\ \hline
     $\text{loss}_{F,t}$ & Generic loss function for neural network $F$ and target label $t$. Takes an input image $x$. \\ \hline
\end{tabular}
\end{table}

\subsection{White box attacks}
In this section, we summarize the most popular white box attacks in the literature.

 \subsubsection{L-BFGS}
 Szegedy et. al. \cite{Szegedy2014} introduced the generation of adversarial examples using a box-constrained L-BFGS (Limited-memory Broyden-Fletcher-Goldfarb-Shanno) algorithm. Given an image $x$ and target label $t$, the proposed method searches for a similar image $x'$ (measured by the $l_2$ distance) that is classified as $t$. The problem is modeled as follows:
 $$ \text{minimize} ||x - x'||^{2}_{2} $$
 $$ \text{such that} \ C(x') = t $$
 $$\ \ \ \ \ \ \ \ \ \ \ \ \ \ \ \ x' \in [0, 1]^n $$
Because finding a solution to this problem can be difficult, they instead solve the following problem:
$$ \text{minimize} \ c \cdot ||x-x'||^2_2 + \text{loss}_{F,t}(x') $$
$$ \text{such that} \ x' \in [0, 1]^n $$
where $\text{loss}_{F,t}$ is any loss function, such as cross-entropy loss.

 \subsubsection{FGSM}
 Goodfellow et. al. \cite{Goodfellow14} introduced the Fast Gradient Sign Method (FGSM) for generating adversarial examples quickly. Given an image $x$, FGSM produces an adversarial $x'$ such that
 $$x' = x + \epsilon \cdot \text{sign}(\Delta\text{loss}_{F,t}(x)) $$
 where $\epsilon$ is chosen to be sufficiently small to avoid detection. 
 
  \textit{Iterative extension.}
  Kurakin et. al. \cite{Kurakin16} extended FGSM to be iterative with the goal of produce adversarial examples that are closer to or more indistinguishable from the original image. This was done by taking multiple steps of size $\alpha$ in the direction of the gradient-sign instead of taking a single step of size $\epsilon$. This iterative process begins with 
  $$ x'_0 = 0 $$
  and each step $x'_i$ is as follows:
  $$x'_i = x'_{i-1} - \text{clip}_\epsilon(\alpha \cdot \text{sign}(\Delta\text{loss}_{F,t}(x'_{i-1})))$$
  
 \subsubsection{JSMA}
 Papernot et. al. \cite{Papernot15} introduced the Jacobian-based Saliency Map Attack (JSMA) for generating adversarial examples based on understanding the mapping between input features and the computed output label. For a neural network $F$ and a target label $t$, this method uses the gradient $\Delta Z_F(x)_t$ to compute a saliency map to model each input pixel's impact on the the computed label of $x$, $C(x)$. Using the saliency map, the most important pixel is modified. This is repeated until an adversarial example is generated or until a set threshold for pixel modifications is reached. For the exact formulation of the saliency map, we recommend reading the original paper.
 
 \subsubsection{DeepFool.}
 Moosavi-Dezfooli et. al. \cite{Moosavidezfooli15} acknowledged the difficulty of understanding the decision making process of a DNN and instead approached a simpler version of the problem. The proposed method assumes a simplified model where the target neural network is completely linear with linearly separable classes. The proposed attack, DeepFool, generates an adversarial example for this simplified problem. This is then repeated until an adversarial example is found for the simpler model that transfers to the non-simplified problem. Similar to our suggestion for JSMA, the full formulation is best found in the original paper.
 
 \subsubsection{C\&W}
 \cite{Carlini16} proposed a new approach to generating adversarial examples, popularly called the Carlini-Wagner (C\&W) method or attack. The C\&W attack finds a small perturbation $\delta$ by solving
    \begin{equation}
    \label{eq:cw}
        \min_{\delta \in \Re^n} \ \ \ ||\delta||_p + c \cdot f(x + \delta)
    \end{equation}
    \[
        \mbox{s.t.} \ \ \ \ \ \ x +  \delta \in [0, 1]^n
    \]
    
    where $f$ is an objective function that leads to $C(x) = t$ and $||\cdot||_p$ is the $l_p$ norm. To solve the box constrain problem, instead of optimizing over $\delta$ as in Equation \ref{eq:cw}, the authors proposed omtimizting over $\omega$ by setting
    \begin{equation}
        \delta_i = \frac{1}{2}(\tanh(\omega_i) + 1) - x_i
    \end{equation}
    
    There are three proposed attacks, an $l_2$ attack, $_0$ attack, and an $l_\infty$ attack, but we only describe the $l_2$ attack as the authors reported it to be the strongest.
    
    The $l_2$ attack attempts to minimize the distortion by searching for an $\omega$ that minimizes
    \begin{equation}
        ||\frac{1}{2}(\tanh (\omega) + 1) - x||^2_2 + c\cdot f(\frac{1}{2}(\tanh (\omega) + 1)
    \end{equation}
    where $f$ is the function
    \begin{equation}
        f(x') = \max(\max\{Z(x')_i : i \ne t\} - Z(x')_t, -\kappa)
    \end{equation}
    where $t$ is a chosen target class, $\kappa$ is a parameter to adjust misclassification confidence, $x'$ is the adversarial example, and $Z(\cdot)$ is the output of the classifier.
   
 \subsubsection{BPDA}
 More recently, Athalye et. al. introduced the Backward Pass Differentiable Approximation (BPDA) attack in response to defenses that obfuscated the target neural network's gradients. Simply put, the gradient of $F(x)$ is approximated by finding a differentiable approximation $G(x)$. To estimate the gradient, the adversary does backpropagation with $F(x)$ during the forward pass but with $G(x)$ during the backward pass.
 
 \subsubsection{Deep representation manipulation.}
 Sabour et. al. \cite{Sabour15} proposed generating adversarial examples guided by the internal representation of a deep neural network. The proposed attack uses a guide image $x_g$ and its internal representation, $Ir_F(x_g)$, with respect to $F(\cdot, \theta)$, and applies perturbations to a different sample image $x$ such that $Ir_F(x) \sim Ir_F(x_g)$. The actual perturbations to push $Ir_F(x)$ towards $Ir_F(x_g)$ are found using L-BFGS from \cite{Szegedy2014}.
 
\subsection{Black box attacks}
In this section, we summarize the most popular black box attacks in the literature.
\subsubsection{Finite difference estimation.}
In finite difference estimation black box attacks, the adversary has access to the class probabilities output by the target model $F$. In this section, we describe black box attacks that take advantage of this information.

  \textit{ZOO}.
  The Zeroth Order Optimization based black box (ZOO) attack \cite{Chen17} is a method for creating adversarial examples that uses finite difference estimates of the gradients. ZOO adopts an iterative optimization approach, similar to the C\&W attack \cite{Carlini16}. The attack begins with a correctly classified input $x$. Then the adversary defines an adversarial loss function that scores perturbations $\delta$ applied to the input and optimizes over the adversarial loss function using the estimated gradients to find $\delta^*$ that creates an adversarial example. ZOO uses "zeroth order stochastic coordinate descent" to optimize the input with respect to the adversarial loss directly.
 
  \textit{Limited Queries and Information.}
  A similar finite difference based approach is adopted by Ilyas et. al. \cite{Ilyas18} in a query limited (QL) setting. Like ZOO, the QL attack estimates the adversarial loss's gradients using a finite difference estimation. However, the QL attack attempts to minimize the number of queries needed by the adversary to estimate the gradients using a search distribution. The QL attack updates the adversarial example using Projected Gradient Descent (PGD) as outlined in Madry et. al. \cite{Madry17} based on the gradient estimates, which are evaluated using Natural Evolutionary Strategies (NES) \cite{Wierstra14}.
  
  This method was later extended upon by the original authors by using a bandit optimization-based algorithm (BAND) with \textit{gradient priors} \cite{Ilyas18b}. The authors defined two types of gradient priors: (1) time-dependent priors or information from successive gradient estimations and (2) data-dependent priors or information from the structure of the inputs. BAND achieves similar success rates to the QL attack using 90\% less queries.
 
 \subsubsection{Transferability}
  Transferability of adversarial examples was introduced by Papernot et. al. \cite{Papernot2016}. This property can be used to mount a black box attack on a target DNN by instead attacking a substitute model and \textit{transferring} the resulting adversarial examples to the target black box model. This substitute model is generally trained for the same classification problem as the target model. Because this model is trained by the adversary, the adversary can use any of the white box attacks described above to generate an adversarial example. 
 
  \textit{Universal perturbations.}
  Moosavi-Dezfooli et. al. \cite{Moosavidezfooli16} also demonstrated this transferability property through a universal adversarial attack. In their experimentation, the authors used DeepFool \cite{Moosavidezfooli15} to generate adversarial examples and found that an attack on model $A$ transfers to a model $B$ trained for the same classification task. The authors also provide a formulation and explanation for the transferability and universality of perturbations.
 
  \textit{Substitution attack with Jacobian-based dataset augmentation.}
  This transferability property was extended by Papernot et. al. \cite{Papernot2016} in a substitution attack. Instead of solely relying on transferability, the authors proposed training a substitute model using data labeled by the target model. Additionally, the authors introduced Jacobian-based Dataset Augmentation, which uses a similar idea of a saliency map in JSMA \cite{Papernot15}, to approximate the target model's decision boundaries. The goal of this attack is to increase the probability of transferable adversarial examples by pushing the substitute model to learn the same decision boundaries as the target model. Using Jacobian-based dataset augmentation decreases the number of queries to the target model during the training phase of the substitute model because the training dataset is \textit{augmented} to heavily weigh exploring decision boundaries of the model. They experimentally showed their attack successfully evading known image classifiers as well as online black box services.
 
 \textit{Generative Adversarial Network (GAN) approach.}
 Zhao et. al. \cite{Zhao17} proposed using a generative adversarial network GAN \cite{Goodfellow14b} to generate adversarial examples. The proposed method trained a GAN with a generator $G$ that maps latent space vectors $Z$ to natural image samples $X$. An \textit{inverter} $I$ is separately trained to map natural image samples $X$ to latent space vectors $Z$, or invert the mapping of $G$. Instead of generating and applying perturbations in the $X$ space, the latent space vector $z_i = I(x_i)$ is perturbed to produce $\hat{z}_i$. An adversarial example in this model is a $\hat{z}_i$ such that $F(G(\hat{z}_i)) \ne F(x)$. This attack relies on the transferability property as the GAN is trained separately from the target model. However, the GAN can be trained using data labeled using the target model, depending on the adversary's knowledge, to incorporate the target model into the attack. Similar approaches using GANs have been proposed by Wang et. al. \cite{Wang19} and Xiao et. al. \cite{Xiao19}.
 
  \textit{$\aleph$-Attack.}
  Motivated by NES \cite{Wierstra14}, Li et. al. \cite{Li19} introduced a powerful black box attack called $\aleph$-Attack. Unlike other existing attacks that optimize perturbations specifically for an input $x$, the proposed attack estimates a probability density distribution centered around the input $x$ such that a sample drawn from the estimated distribution is an adversarial example. $\aleph$-Attack essentially learns the distribution of adversarial examples and samples that distribution to generate a black box attack. The authors experimentally showed that $\aleph$-Attack was more effective than the BPDA attack \cite{Athalye18} against ML models trained on the CIFAR10 and Imagenet datasets.
 
\end{document}